\documentclass[twocolumn,showpacs,superscriptaddress,prl]{revtex4-1}
\usepackage{amssymb}
\usepackage{amsmath}
\usepackage[dvips]{graphicx}
\usepackage{dcolumn}
\usepackage{bm}
\usepackage{color}

\begin{document}

\title{Electric field-induced Skyrmion distortion and giant lattice rotation in the magnetoelectric insulator Cu$_{2}$OSeO$_{3}$}

\author{J.\,S.\,White}\email{jonathan.white@psi.ch}
\affiliation{Laboratory for Quantum Magnetism, Ecole Polytechnique F\'{e}d\'{e}rale de Lausanne (EPFL), CH-1015 Lausanne, Switzerland}
\affiliation{Laboratory for Neutron Scattering, Paul Scherrer Institut, CH-5232 Villigen, Switzerland}

\author{K.\,Pr\v{s}a}
\author{P.\,Huang}
\author{A.\,A.\,Omrani}
\affiliation{Laboratory for Quantum Magnetism, Ecole Polytechnique F\'{e}d\'{e}rale de Lausanne (EPFL), CH-1015 Lausanne, Switzerland}

\author{I.\,\v{Z}ivkovi\'{c}}
\affiliation{Institute of Physics, Bijeni\v{c}ka 46, HR-10000 Zagreb, Croatia}

\author{M.\,Bartkowiak}
\affiliation{Laboratory for Developments and Methods, Paul Scherrer Institut, CH-5232 Villigen, Switzerland}

\author{H.\,Berger}
\author{A.\,Magrez}
\affiliation{Crystal Growth Facility, Ecole Polytechnique F\'{e}d\'{e}rale de Lausanne (EPFL), CH-1015 Lausanne, Switzerland}

\author{J.\,L.\,Gavilano}
\author{G.\,Nagy}
\affiliation{Laboratory for Neutron Scattering, Paul Scherrer Institut, CH-5232 Villigen, Switzerland}

\author{J.\,Zang}\email{jiadongzang@gmail.com}
\affiliation{Department of Physics and Astronomy, Johns Hopkins University, Baltimore, Maryland 21218, USA}

\author{H.\,M.\,R\o nnow}\email{henrik.ronnow@epfl.ch}
\affiliation{Laboratory for Quantum Magnetism, Ecole Polytechnique F\'{e}d\'{e}rale de Lausanne (EPFL), CH-1015 Lausanne, Switzerland}

\date{\today}

\begin{abstract}
Uniquely in Cu$_{2}$OSeO$_{3}$, the Skyrmions, which are topologically protected magnetic spin vortex-like objects, display a magnetoelectric coupling and can be manipulated by externally applied electric ($E$) fields. Here we explore the $E$-field coupling to the magnetoelectric Skyrmion lattice phase, and study the response using neutron scattering. Giant $E$-field induced rotations of the Skyrmion lattice are achieved that span a range of $\sim$25$^{\circ}$. Supporting calculations show that an $E$-field-induced Skyrmion distortion lies behind the lattice rotation. Overall we present a new approach to Skyrmion control that makes no use of spin-transfer torques due to currents of either electrons or magnons.
\end{abstract}

\maketitle

Skyrmions are magnetic spin vortex-like objects that can be stabilised in chiral magnets with Dzyaloshinskii-Moriya (DM) interactions. Due to their topological property~\citep{Muh09,Nag13}, nanometric size ($\sim $10-100~nm)~\citep{Nag13}, observation close to room temperature ($T$)~\citep{Leb89,Yu11}, and conduction electron-driven motion~\citep{Jon10,Yu12,Fer13}, Skyrmions are outstanding candidate components for new spin-based applications.

Materials that display Skyrmions range from itinerant MnSi~\citep{Muh09,Neu09}, Mn$_{1-x}$Fe$_{x}$(Si,Ge)~\citep{Gri09,Shi13} and FeGe~\citep{Yu11}, to semi-conducting Fe$_{1-x}$Co$_{x}$Si~\citep{Mun10}, and insulating Cu$_{2}$OSeO$_{3}$~\citep{Sek12,Ada12}. All have the chiral-cubic space group $P2_{1}3$, a weak magneto-crystalline anisotropy, and common phase diagrams with a helimagnetic groundstate. Despite these similarities, the diverse transport properties lead to material specific mechanisms for Skyrmion manipulation and the associated dynamics. In the well-studied itinerant compounds, spin-transfer torques (STTs) exerted by the conduction electrons of an ultra-low current density, $j$$\lesssim$10$^{6}$A.m$^{-2}$ drive the Skyrmion motion~\citep{Jon10,Eve11,Zan11,Eve12,Sch12,Lin13,Iwa13}. More generally, in both MnSi and insulating CuO$_{2}$SeO$_{3}$, Skyrmion lattice (SkL) rotations are observed to be driven by STTs exerted by the magnon currents induced by a thermal gradient~\citep{Moc14}. Even though electric currents and thermal gradients have been established to generate Skyrmion motion, it remains vital to find new control mechanisms which may lead to further efficient Skyrmion-based functionalities.

In the insulating SkL host compounds, the chiral lattice promotes a magnetoelectric (ME) coupling between electric ($E$) and magnetic orders which can be expected to lie at the heart of new Skyrmion control paradigms. The use of ME coupling for Skyrmion manipulation is also attractive for applications since losses due to Joule heating are negligible. Presently however, open questions remain concerning the basic understanding of how an applied $E$-field can manipulate the Skyrmion spin texture. To address this issue, we have used small-angle neutron scattering (SANS) to study the giant $E$-field-induced SkL rotations generated in a bulk sample of ME Cu$_{2}$OSeO$_{3}$. Surprisingly, the rotations saturate at an angle dependent on both the size and sign of the $E$-field. With supporting calculations we explain our observations, and show that an $E$-field-induced Skyrmion distortion leads to the observed rotations. This amounts to a new approach for Skyrmion control that does not require STTs.

In Cu$_{2}$OSeO$_{3}$ the ME coupling exists in all magnetic phases~ \citep{Bos08,Bel10,Bel12,Sek12,Sek12b,Mai12,Ziv12,Yan12,Omr14}, and is generated by the $d$-$p$ hybridization mechanism~\citep{Sek12,Sek12b,Jia07,Ari07}. This mechanism dictates a particular ME coupling anisotropy; for a magnetic field $\mu_{0}H$$\parallel$$[110]$ or $[111]$, an electric polarisation $P$ emerges $\parallel$$[001]$ or $[111]$, respectively~\citep{Sek12b}. In our experiments we chose $E$$\parallel$$[111]$ (which corresponds to a negative applied voltage) or $\parallel$$[\bar{1}\bar{1}\bar{1}]$ (positive voltage). The direction of $\mu_{0}H$ could be chosen anywhere in a horizontal plane defined by the $[1\bar{1}0]$ and $[111]$ cubic axes of the sample. This setup allowed SANS studies of two main geometries where finite ME coupling is expected, both $E$$\parallel$$\mu_{0}H$$\parallel$$[111]$, and $E$$\parallel$$[111]$ with $\mu_{0}H$$\parallel$$[1\bar{1}0]$. For each geometry, $\mu_{0}H$ was always approximately parallel to the incident neutron beam, thus allowing SANS imaging in both the SkL and zero-field phases. The SANS experiments were conducted at SINQ, Paul Scherrer Insitut, Switzerland. Full details of our experiments are found in Ref.~\citep{Sup}.

\begin{figure}[tbp]
\includegraphics[width=0.46\textwidth]{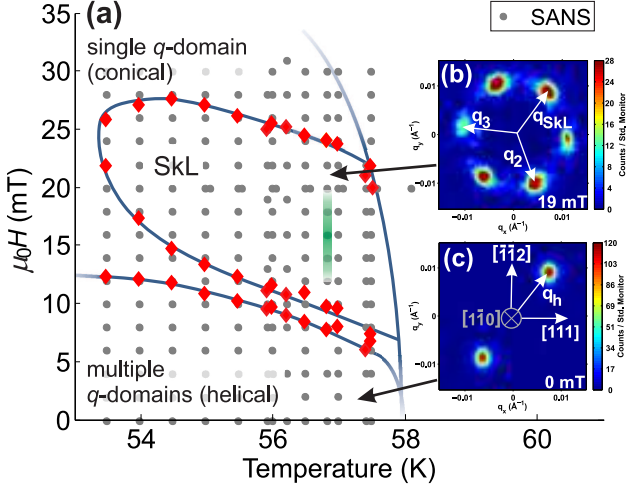}
\caption{(a) The high-$T$ portion of the magnetic phase diagram of Cu$_{2}$OSeO$_{3}$ for $\mu_{0}H$$\parallel$$[1\bar{1}0]$. Gray points denote where SANS measurements were done, and red diamonds the estimated locations of the phase boundaries. Similar phase diagrams are reported elsewhere (notwithstanding quantitative differences expected due to demagnetization effects)~\citep{Ada12,Sek12b,Omr14}. In green we indicate the peak-to-peak range of oscillating $\mu_{0}H$ where a dc $E$-field was also applied (see text for details). (b) and (c) respectively show typical SANS images from the SkL phase at 19~mT, and the zero-field helical phase.}
\label{fig:phase_diagram}
\end{figure}

Fig.~\ref{fig:phase_diagram}(a) shows the explored portion of the magnetic phase diagram for $\mu_{0}H$$\parallel$$[1\bar{1}0$], and for our 26~mg single crystal of Cu$_{2}$OSeO$_{3}$. In the zero-field helical phase, we observe the expected SANS pattern with a single wavevector \textbf{q}$_{\mathrm{h}}$$\parallel$$[001]$ (note that both $\pm$\textbf{q} each give a Bragg spot) [Fig.~\ref{fig:phase_diagram}(c)]. By applying $\mu_{0}H$$\parallel$$[1\bar{1}0]=19$~mT at 56.8~K, the SANS pattern transforms into the six-fold symmetric one expected for the triple-\textbf{q} SkL, also with a wavevector \textbf{q}$_{\mathrm{SkL}}$$\parallel$$[001]$, and two further \textbf{q}-vectors at angles relative to \textbf{q}$_{\mathrm{SkL}}$ of $\pm2\pi/3$ [Fig.~\ref{fig:phase_diagram}(b)]. Consistent with previous work~\citep{Whi12}, applying a dc $E$-field within the SkL state does not discernibly alter the SkL orientation. Here we discover that by additionally oscillating $\mu_{0}H$ weakly around its mean value, a SkL rotation is generated that saturates at an angle dependent solely on the $E$-field.


\begin{figure}[tbp]
\includegraphics[width=0.46\textwidth]{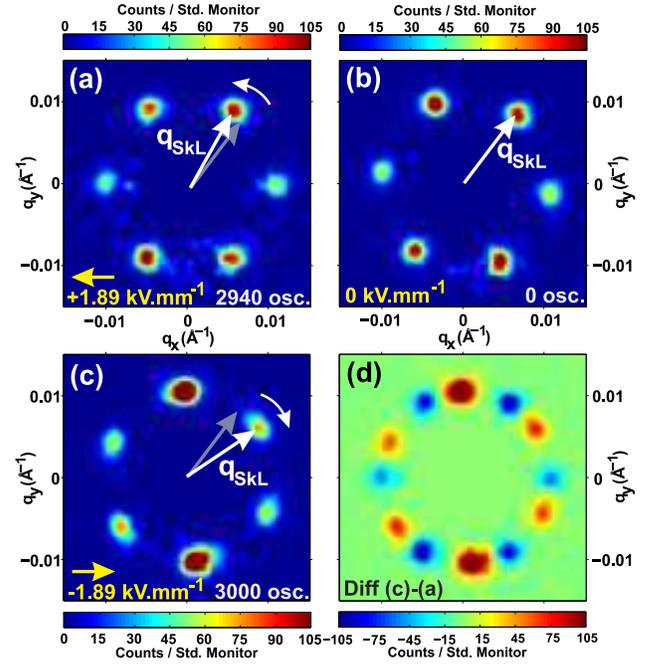}
\caption{SANS diffraction patterns from the SkL with the sample under (a) $E$=+1.89~kV.mm$^{-1}$ and after 2940 $\mu_{0}H$ oscillations, (b) $E$=0, and no $\mu_{0}H$ oscillations, and (c) $E$=-1.89~kV.mm$^{-1}$ and after 3000 $\mu_{0}H$ oscillations. All measurements were done at $T=$~56.8~K, and with a static $\mu_{0}H$$\parallel$$[1\bar{1}0$]=16~mT. In (a)-(c) the diffraction spots exhibit different intensities since they move at different speeds through the Bragg condition. (d) shows the foreground data of panel (a), subtracted from those of panel (c). All images correspond to views of the detector as seen from the sample.}
\label{fig:SANS_patterns}
\end{figure}

Fig.~\ref{fig:SANS_patterns} summarizes the $E$-field-induced SkL rotations generated in Cu$_{2}$OSeO$_{3}$. Fig.~\ref{fig:SANS_patterns}(b) shows the SkL orientation with \textbf{q}$_{\mathrm{SkL}}$$\parallel$$[001]$ after an initial zero-field cool from 70~K to 56.8~K, followed by ramping $\mu_{0}H$$\parallel$$[1\bar{1}0]$ to 16~mT. By sequentially applying an $E$-field=+1.89~kV.mm$^{-1}$, and then 2940 triangular $\mu_{0}H$ oscillations around the average 16~mT field, the SkL orientation shown in Fig.~\ref{fig:SANS_patterns}(a) is obtained. Here the $\mu_{0}H$ oscillations had a frequency $f=0.05$~Hz, and an amplitude $\Delta\mu_{0}H\pm$4~mT (8~mT peak-to-peak) chosen so as to always remain within the SkL phase - see the green line in Fig.~\ref{fig:phase_diagram}(a). From Fig.~\ref{fig:SANS_patterns}(a) we find that the SkL orientation has rotated counter-clockwise relative to that seen in Fig.~\ref{fig:SANS_patterns}(b). Repeating the overall procedure except with a reversed $E$-field of -1.89~kV.mm$^{-1}$, Fig.~\ref{fig:SANS_patterns}(c) shows the SkL to have rotated clockwise after 3000 oscillations. In Fig.~\ref{fig:SANS_patterns}(d) we emphasize the contrasting SkL orientations obtained for $E$-fields of opposite direction.

\begin{figure}
\includegraphics[width=0.46\textwidth]{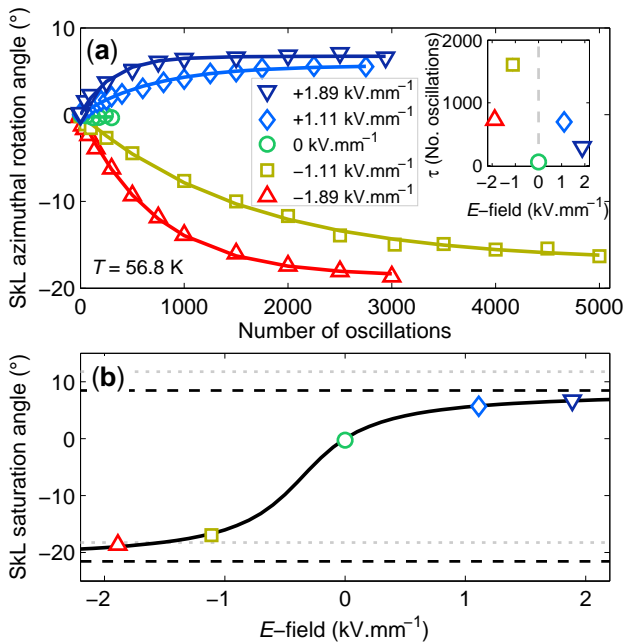}
\caption{(a) The average SkL rotation angle as functions of both $\mu_{0}H$ oscillation number and $E$-field. At each datapoint, the oscillations were stopped and the SANS measurement
done in a static $\mu_{0}H$=16~mT. A negative (positive) angle corresponds to a clockwise (counter-clockwise) rotation of \textbf{q}$_{\mathrm{SkL}}$ as seen in Fig.~\ref{fig:SANS_patterns}. Solid lines are fits using the relaxation model described in the text. From the fits, we obtain the $E$-field-dependence of $\tau(E)$ (inset), and the rotation saturation angle shown in (b). In (b), the solid black line is a fit of the data as described in the text. The fit parameters define the range of possible SkL orientations that lie between the dashed black lines. Dotted gray lines indicate the range of orientations expected directly from the theory.}
\label{fig:wiggle_curves}
\end{figure}

Fig.~\ref{fig:wiggle_curves}(a) shows the average SkL rotation angle as a function of $\mu_{0}H$ oscillation number for several $E$-fields. Three salient results arise from the analysis. First, for the explored $E$-field range the rotations span a giant range of $\sim$25$^{\circ}$. Second, the rotations are asymmetric with $E$-field, with larger rotations achieved for negative $E$-fields. Third, at each $E$-field the rotation angle saturates at large oscillation number. To extract quantitative information, we fitted the simple relaxation model $\Phi(n,E)$=$\Phi_{0}(E)\left(1-\text{exp}(-n/\tau(E))\right)$ to the data at each $E$-field. Here $\Phi(n,E)$ is the average SkL rotation angle relative to 0$^{\circ}$ (with \textbf{q}$_{\mathrm{SkL}}$$\parallel$$[001]$), and $n$ is the number of $\mu_{0}H$ oscillations. $\Phi_{0}(E)$ is the saturation rotation angle, and $\tau(E)$ a relaxation constant. From the fits to each curve seen in Fig.~\ref{fig:wiggle_curves}(a), we obtain the $E$-field-dependence of both $\tau(E)$ [inset Fig.~\ref{fig:wiggle_curves}(a)] and $\Phi_{0}(E)$ [Fig.~\ref{fig:wiggle_curves}(b)]. Preliminary data show $\tau(E)$ to depend strongly on $\Delta\mu_{0}H$, and weakly on $f$ for the low frequency range 0.02-0.50~Hz. Overall, our approach gives novel microscopic insight into the quasi-dynamic response of the SkL orientation.


Next we consider unexpected effects that could explain our observations. First, with $E$=0, various $T$- and $\mu_{0}H$-dependent measurements show just the single SkL orientation with \textbf{q}$_{\mathrm{SkL}}$$\parallel $$[001]$ to exist in the explored portion of the SkL phase [Fig.~\ref{fig:phase_diagram}(a)]. No evidence is found in our sample for a $\mu_{0}H$- and/or $T$-driven SkL realignment within the SkL phase like that reported in Ref.~\citep{Sek12c}, not even after applying $\mu_{0}H$ oscillations. The reason for such a difference in observations is unclear, though it may reflect slight variations in the microscopic properties between samples. Second, when oscillating $\mu_{0}H$ and obtaining the curves shown in Fig.~\ref{fig:wiggle_curves}(a), we chose our experimental conditions so as to never cross the SkL phase boundary. This was done to avoid any effect due to $E$-field poling which was shown previously may influence the SkL orientation~\citep{Whi12}. Third, any residual leakage currents under applied $E$-fields were always $\lesssim$10$^{-2}$A.m$^{-2}$, and insufficient to induce Skyrmion motion. Fourth, our experiments were designed for minimized gradients across the sample. With $E=$~0, no rotation was detected at nominally constant sample $T$. This indicates that control mechanisms reliant solely on a thermal gradient can be discounted~\citep{Kon13,Moc14}. A persistent SkL rotation could be imagined under a spatial $E$-field gradient~\citep{Liu13}, though this is inconsistent with the observed saturation of the SkL orientation.

Instead, we develop an analytic model to show that an intrinsic ME effect lies behind the rotations. We consider a spin Hamiltonian $\mathcal{H}$ that approximates the full microscopic Hamiltonian for Cu$_{2}$OSeO$_{3}$ by a simplified one which contains a single effective spin $\mathbf{S}$ per cubic unit cell~\citep{Yi09,Liu13,Sup}. The leading term in $\mathcal{H}$ is $\mathcal{H}_{HDM}$=$J(\nabla \mathbf{S})^{2}+D\mathbf{S}\cdot(\nabla \times \mathbf{S})-\mathbf{h}\cdot \mathbf{S}$, where $J$, $D$ are the coefficients of the Heisenberg and DM interactions respectively, and $\mathbf{h}$ is the magnetic field. For convenience, the natural coordinate system is rotated so that the new $\hat{y}$ axis coincides with [001] in the original frame, and the new $\hat{z}$ axis coincides with [1\={1}0], which is parallel to $\mu_{0}H$. In the absence of $\mathbf{h}$, the competition between the Heisenberg and DM interactions leads to a helical groundstate with spin configuration $\mathbf{S}(\bm{r})$=$\mathbf{S}(\mathbf{q})e^{i\bm{q}\cdot\bm{r}}+\mathbf{S}(-\mathbf{q})e^{-i\bm{q}\cdot \bm{r}}$, and where $\mathbf{S}(\mathbf{q})$=$1/\sqrt{2}(-i\hat{q}^{y},i\hat{q}^{x},1)$ is the Fourier amplitude of the helix at propagation vector $\mathbf{q}$. At finite $\mathbf{h}$ the SkL can be stabilized, and the spin configuration can be written as a superposition of three proper helices~\citep{Muh09}:
\begin{equation}
\mathbf{S}(\mathbf{r})=m_{H}\sum_{i=1}^{3}\left( \mathbf{S}(\mathbf{q}_{i})e^{i\bm{q}_{i}\cdot \bm{r}}+\mathbf{S}(-\mathbf{q}_{i})e^{-i\bm{q}_{i}\cdot \bm{r}}\right) +m_{F}\hat{z}.
\label{Eq:2}
\end{equation}
Here propagation vectors $\textbf{q}_{i}$ ($i$=1,2,3) lie in the $\hat{x}$-$\hat{y}$ plane, and have azimuthal angles of $\phi$, $\phi+2\pi/3$, and $\phi-2\pi/3$, respectively. The angle $\phi$ thus defines the SkL orientation. $m_{H}$ and $m_{F}$ are the components of the helices and the net magnetization, respectively. They satisfy the constraint $1$=$\langle
S^{2}(\mathbf{r})\rangle$=$6m_{H}^{2}+m_{F}^{2}$, so that the spins in the lattice all have unit length on average.

Next we examine the magneto-crystalline anisotropy term, $\mathcal{H}_{A}$ which is constructed for the cubic lattice using a symmetry analysis that employs Neumann's principle~\citep{Sup}. Substitution of Eq.~\ref{Eq:2} into each of the second- and fourth-order spin anisotropy terms yield constants independent of $\phi$. A $\phi$-dependent energy is obtained however from the sixth-order term;
\begin{equation}
\mathcal{E}_{A}^{(6)}=B(5/512)m_{H}^{6}\textrm{cos}(6\phi)
\label{Eq:2p5}
\end{equation}
where $B$ represents the amplitude of sixth-order spin anisotropy. Minimizing $\mathcal{E}_{A}^{(6)}$ yields solutions $\phi$=0 when $B$$<$0, and $\phi$=$\pi/6$ when $B$$>$0. For $E$=0, we observe that the SkL has a propagation vector parallel to $[001]$. This direction corresponds to the $\hat{y}$-axis in the rotated model frame, so $\phi$=$\pi/6$. Therefore, $B$$>$0 ensures the correct minimization of $\mathcal{E}_{A}^{(6)}$ that is consistent with our experiments, though note that the following theory is generally applicable for either sign of $B$.

For finite electric field, $E$ a ME coupling term, $\mathcal{H}_{\mathrm{ME}}$ needs to be included in $\mathcal{H}$. In the rotated coordinate frame, $\mathcal{H}_{\mathrm{ME}}=\alpha E(S_{x}^{2}+\sqrt{2}S_{x}S_{y}-S_{z}^{2})/2$, where $\alpha$ represents the strength of the ME coupling. To zeroth order, substitution of Eq.~\ref{Eq:2} into $\mathcal{H}_{\mathrm{ME}}$ again yields merely a constant that is $\phi$-independent. Therefore we have to consider higher-order effects in $E$.

To this end, the first-order perturbation in $E$ is employed~\citep{Sup}. We find that the influence of $E$ is to distort the SkL, since $\mathbf{S}(\mathbf{q})$ for each helix acquires components of an anti-screw helix with $\mathbf{S}'(\mathbf{q})=1/\sqrt{2}(i\hat{q}^{y},-i\hat{q}^{x},1)$, and a third orthogonal spiral with $\mathbf{S}''(\mathbf{q})=(\hat{q}^{x},\hat{q}^{y},0)$~\citep{Sup}. The perturbation is treated up to first-order in $\beta(E)$=$\alpha E/Dk_{0}$, where $k_{0}$=$D/2J$ is the length of the SkL propagation vector. Using estimates $J$=50~K, $D$=3~K, and $\alpha$=$10^{-33}$J/(V/m)~\citep{Omr14}, $\beta(E)$$\sim$10$^{-3}$ for $E$=1~kV.mm$^{-1}$. Thus, treating the perturbation to first-order is well justified, and the relative changes in the overall Fourier components are small. For $|\beta(E)|$=0.001, we calculate $|\mathbf{S}(\mathbf{q})|^{2}$, which is proportional to the measured SANS intensity at wavevector $\textbf{q}$, to vary by just $\sim$10$^{-3}$$\%$ for each SkL propagation vector. This challenges a reliable detection of the distortion by SANS.

Nevertheless, on substitution of the distorted Skyrmion configuration into $\mathcal{H}_{A}$, a $\phi$-dependence that can influence the SkL alignment now appears at fourth-order in spin operator. Its energy is given by
\begin{equation}
\mathcal{E}_{A}^{(4)^{\prime}}=\frac{810\sqrt{2}\alpha m_{H}^{4}E}{512Dk_{0}}A\left( \text{sin}6\phi +\frac{\sqrt{2}}{4}\text{cos}6\phi \right).
\label{Eq:3}
\end{equation}
Here $A$ represents the amplitude of fourth-order anisotropy. Combining $\mathcal{E}_{A}^{(4)^{\prime }}$ with the unperturbed energy $\mathcal{E}_{A}^{(6)}$ gives an overall anisotropy energy that describes a balance between magneto-crystalline and $E$-field-induced anisotropies:
\begin{equation}
\mathcal{E}_{A}=\frac{3m_{H}^{4}}{512}\left( \left( \xi+\frac{\sqrt{2}}{4}\lambda \right) ^{2}+\lambda ^{2}\right)^{\frac{1}{2}}\text{cos}6\left(\phi+\theta_{\rm E}\right)
\label{Eq:5}
\end{equation}
where $\lambda=270\sqrt{2}A\alpha E/Dk_{0}$ and $\xi=5m_{H}^{2}B$. Eq.~\ref{Eq:5} shows that under finite $E$ the ground state minimum is rotated away from the $E=0$ case by an azimuthal angle $\theta_{\rm E}$:
\begin{equation}
\theta_{\rm E}=\frac{1}{6}\text{arctan}\left(\frac{\lambda}{\xi + C\lambda}\right),
\label{Eq:6}
\end{equation}
where $C$ is expected to be $\sqrt{2}/4$~\citep{Sup}. In the limit $B$=0, evaluation of Eq.~\ref{Eq:6} yields two limiting orientations that are separated by 30$^{\circ}$ - see the dotted gray lines in Fig.~\ref{fig:wiggle_curves}(b). The size of asymmetry of these orientations around 0$^{\circ}$ is determined by coefficient $C$.

Fig.~\ref{fig:wiggle_curves}(b) shows that the data of rotation saturation angle versus $E$-field indeed follow the expected tangential form of $\theta_{\rm E}$ [Eq.~\ref{Eq:6}]. However, the observed values span a range that is shifted slightly compared with that expected. A full description of the data is achieved upon fitting with the recast form of Eq.~\ref{Eq:6}: $\theta_{\rm E}$=$\frac{1}{6}\text{arctan}(1/(K/E + C))$, where $K$=$(5m_{\mathrm{H}}^{2}Dk_{0})/(270\sqrt{2}\alpha)(B/A)$. From the fit - the solid black line in Fig.~\ref{fig:wiggle_curves}(b) - we obtain $K$=0.71(1)~kV.mm$^{-1}$ and $C$=0.82(1). Using the value of $K$, and equating $m_{\mathrm{H}}$=1.1$\mu_{\mathrm{B}}$/unit cell as estimated from the data in Ref.~\onlinecite{Omr14}, we find $B/A\sim10^{-3}$ which is self-consistent with our energy considerations. The fitted value for $C$ is larger than $\sqrt{2}/4\simeq 0.35$ expected from the theory. This discrepancy could be resolved by including further higher-order corrections, such as those obtained from a perturbative treatment of the sixth-order spin anisotropy. Such corrections would leave the general form of Eq.~\ref{Eq:6} unchanged, yet modify the expected value of $C$. Nevertheless, the robustness of the approach is seen when evaluating the model for other $\mu_{0}H$ and $E$-field configurations. Importantly, no rotations are expected from our theory for the $E$$\parallel$$\mu_{0}H$$\parallel $$[111]$ geometry~\citep{Sup}, which is consistent with our SANS measurements (data not shown).

By both experiment and theory we have shown that ME coupling allows control over the preferred SkL orientation in Cu$_{2}$OSeO$_{3}$. This is achieved without requiring a macroscopic spatial gradient across the sample, such as the $T$ gradients applied previously~\citep{Jon10,Eve12,Moc14}. Nonetheless, we can not rule out that any finite spatial gradients, be they in $T$ or in $E$-field~\citep{Liu13}, play a role in the rotational dynamics. Indeed, a full parametrization of the insulating SkL dynamics in this model system await further studies. Instead, in our experiments we have explored a regime where Skyrmion pinning dominates over the rotational torque, since simply applying an $E$-field is insufficient to drive the rotation of the equilibrium SkL. Only by driving the system out of equilibrium with $\mu_{0}H$ oscillations is the pinning overcome, and the rotation initiated. Concomitantly, it can be expected that when the $\mu_{0}H$ oscillations are stopped, pinning preserves an achieved SkL orientation, even after further removal of the $E$-field. This phenomenon, which should exist also at room temperature, could be exploited as an information storage scheme.

To summarize our study, we have shown that by means of ME coupling $E$-fields can control SkL rotations in a bulk single crystal of Cu$_{2}$OSeO$_{3}$. With supporting calculations, we demonstrate that the rotations arise due to an $E$-field-induced Skyrmion distortion. Our study presents a new Skyrmion manipulation concept which, unlike all other reported approaches, does not require STTs due to currents of either conduction electrons or magnons.

Financial support from the Swiss National science Foundation, the European Research Council grant CONQUEST, MaNEP and the Indo-Swiss Joint Research Project programme is gratefully acknowledged. We thank D.~Mazzone and U.~Gasser for support with the neutron experiments performed at the Swiss Spallation Neutron Source (SINQ), Paul Scherrer Institut, Switzerland. IZ acknowledges financial support from the Croatian Science Foundation Project No. 02.05/33. JZ is supported by the Theoretical Interdisciplinary Physics and Astrophysics Center and by the U.S. Department of Energy, Office of Basic Energy Sciences, Division of Materials Sciences and Engineering under Award DEFG02-08ER46544.

\bibliography{cu2oseo3}

\end{document}